\documentclass[a4paper]{article}
\usepackage[utf8]{inputenc}
\usepackage{graphicx}
\graphicspath{{./images/}{./} }
\makeatletter
\g@addto@macro\@floatboxreset\centering
\makeatother

\usepackage[style=nature,sorting=none,backend=bibtex,autocite=superscript]{biblatex}
\usepackage[font=small,labelfont=bf]{caption}
\DeclareCaptionLabelSeparator{pipe}{ $\mathbf{|}$ }
\usepackage{hyperref}% Clickable references (and table of contents if applicable)
\usepackage[intlimits]{amsmath}
\usepackage{textcomp} % for \textmu
\usepackage{authblk} % for author affiliation
\usepackage{amssymb} % for \gtrsim

\newcommand{\dorms}{\delta \omega_\mathrm{rms}}
\newcommand{\xzpf}{x_\mathrm{zpf}}
\newcommand{\xth}{x_\mathrm{th}}
\newcommand{\nth}{\overline{n}_\mathrm{th}}

\newcommand{\Pin}{P_\mathrm{in}}
\newcommand{\Plo}{P_\mathrm{LO}}

\newcommand{\oc}{\omega_\mathrm{c}}
\newcommand{\Om}{\Omega_\mathrm{m}}
\newcommand{\varone}[1]{\langle #1 \rangle}
\newcommand{\Hint}{\hat{H}_\mathrm{int}}

\title{Nonlinear cavity optomechanics with nanomechanical thermal fluctuations}
\author[1]{Rick Leijssen}
\author[1]{Giada La Gala}
\author[1]{Lars Freisem}
\author[1]{Juha T. Muhonen}
\author[1,*]{Ewold Verhagen}
\affil[1]{Center for Nanophotonics, AMOLF, Science Park 104, 1098 XG, Amsterdam, The Netherlands}
\affil[*]{corresponding author: verhagen@amolf.nl}
\begin{document}
\begin{refsection}
%\onecolumn
\maketitle
\begin{abstract}
  {\bfseries
The inherently nonlinear interaction between light and motion in cavity optomechanical systems has experimentally been studied in a linearized description in all except highly driven cases.
Here we demonstrate a nanoscale optomechanical system, in which the interaction between light and motion is so large (single-photon cooperativity $C_0 \approx 10^3$) that thermal motion induces optical frequency fluctuations larger than the intrinsic optical linewidth.
The system thereby operates in a fully nonlinear regime, which pronouncedly impacts the optical response, displacement measurement, and radiation pressure backaction.
Experiments show that the apparent optical linewidth is dominated by thermomechanically-induced frequency fluctuations over a wide temperature range.
The nonlinearity induces breakdown of the traditional cavity optomechanical descriptions of thermal displacement measurements.
Moreover, we explore how radiation pressure backaction in this regime affects the mechanical fluctuation spectra.
The strong nonlinearity could serve as a resource to control the motional state of the resonator.
We demonstrate the use of highly nonlinear transduction to perform a quadratic measurement of position while suppressing linear transduction.
}
\end{abstract}

\clearpage
\section{Introduction}
In cavity optomechanics, the interaction between light in an optical cavity and the motion of a mechanical resonator enables sensitive optical readout of displacement, as well as manipulation of the resonator's motion through optical forces\autocite{Aspelmeyer2014}.
This has allowed demonstrations of sideband and feedback cooling of the mechanical resonator near its quantum ground state\autocite{Chan2011,Teufel2011,Verhagen2012,Wilson2015}, squeezing of light\autocite{Brooks2012,SafaviNaeini2013,Purdy2013a} and of the mechanical zero-point fluctuations\autocite{Wollman2015,Pirkkalainen2015,Lecocq2015}, entanglement\autocite{Palomaki2013} and state transfer\autocite{Palomaki2013a} between the optical and mechanical degrees of freedom, as well as detection of radiation pressure shot noise\autocite{Purdy2013,Murch2008} and non-classical correlations\autocite{Riedinger2016,Sudhir2016,Purdy2016,Kampel2016}.
In all of these examples, the coupling between fluctuations of the optical field and the mechanical displacement can be regarded as linear for all intents and purposes.

However, the optomechanical interaction is inherently nonlinear.
Indeed, the cavity optomechanical interaction Hamiltonian reads $\Hint = - \hbar g_0 \hat{a}^\dag \hat{a} \hat{x} / \xzpf$, where $\hat{a}$ is the annihilation operator for the optical cavity field, and $\hat{x}$ is the displacement operator for the mechanical resonator. 
The photon-phonon coupling rate $g_0 = (\partial \oc /  \partial x) \xzpf$ quantifies the change of the cavity frequency $\oc$ due to a displacement the size of the resonator's zero-point fluctuations $\xzpf$.
This interaction Hamiltonian leads to nonlinear behavior, as the equations of motion it generates contain products of two operators. 
The linearized form of the interaction $\hat{H}_\mathrm{int} = - \hbar g_0 \overline{\alpha}(\delta\hat{a}^\dag + \delta\hat{a}) \hat{x} / \xzpf$ does not contain the nonlinear terms, but usually suffices to describe the dynamics of fluctuations\autocite{Aspelmeyer2014}.
This form emerges when the cavity field is written as the sum $\hat{a} = \overline{\alpha} + \delta \hat{a}$ of an average coherent field $\overline{\alpha}$ and fluctuations $\delta\hat{a}$, and the term containing $\delta\hat{a}^\dag \delta\hat{a}\hat{x}$ is neglected by assuming $\delta \hat{a} \ll \overline{\alpha}$.
The linearization is generally valid in the presence of a strong optical drive.
However, the assumption that $\delta \hat{a} \ll \overline{\alpha}$ is not valid if mechanical fluctuations shift the cavity completely in and out of resonance with the optical drive, i.e. when they produce a cavity frequency shift comparable to the optical linewidth $\kappa$.
Then, nonlinear processes become crucially important, and qualitatively different effects can occur.

In the quantum domain, intriguing implications of this nonlinearity are expected in the single-photon strong-coupling regime, when the coupling rate $g_0$ exceeds the optical and mechanical loss rates $\kappa$ and $\Gamma$, respectively. % ($g_0 > \kappa,\Gamma$).
There, quantum-level mechanical fluctuations induce a nonlinear response, creating nonclassical states of both light and motion when the mechanical frequency $\Om$ approaches the optical linewidth as well\autocite{Nunnenkamp2011,Rabl2011,Lemonde2016}.
In the so-called bad-cavity limit ($\kappa > \Om$), the nonlinearity of the interaction provides a useful path towards creating motional quantum states, e.g. through performing quadratic measurements of displacement (proportional to $\hat{x}^2$)\autocite{Doolin2014,Brawley2016,Vanner2013,Khosla2013,Bennett2016}.

In macroscopic or chip-based optomechanical implementations, the breakdown of linearity when $\delta\hat{a}\gtrsim \overline{\alpha}$ has so far only been experimentally relevant for mechanical resonators driven to large amplitude, e.g. through optomechanical parametric amplification.
In that case, nonlinear effects determine the maximum amplitude of optomechanical self-oscillation \autocite{Aspelmeyer2014,Kippenberg2005,Marquardt2006,Vahala2009,Bagheri2011,Krause2015}, and can lead to complex nonlinear dynamical phenomena such as chaos\autocite{Carmon2007,Wu2016,Navarro-Urrios2016}.

Here, we establish and explore the regime where even intrinsic Brownian motion induces cavity frequency fluctuations larger than the optical linewidth.
In this regime, the nonlinear nature of the cavity optomechanical interaction becomes important in all essential phenomena, including optomechanical displacement measurement and radiation pressure backaction.
The regime is defined by $g_0 \sqrt{2 \nth} \gtrsim \kappa$, where $\nth = k_B T / \hbar \Om$ is the average phonon occupancy of the mechanical mode with frequency $\Om$, in thermal equilibrium at a temperature $T$.
It is clear from this condition that any optomechanical system in which the ratio $g_0 / \kappa$ is increased will enter this regime before reaching the single-photon strong-coupling regime, unless the mechanical resonator is pre-cooled to its ground state.
The condition can equivalently be expressed as $C_0\gtrsim\kappa/\gamma$, i.e. the single-photon cooperativity $C_0\equiv 4g_0^2/\kappa\Gamma$ being larger than the ratio of optical decay rate and mechanical thermal decoherence rate $\gamma\equiv\Gamma\nth$.
In our experiments, we follow a strategy of exploiting subwavelength optical confinement\autocite{Leijssen2015} to reach unprecedented single-photon cooperativity around $10^3$; two to three orders of magnitude larger than typical values in nanoscale optomechanical systems to date\autocite{Aspelmeyer2014,Meenehan2015,Schilling2016}, and only comparable with cold-atom implementations\autocite{Murch2008,Brennecke2008}.
In the systems demonstrated here, the apparent optical linewidth is dominated by the transduced thermal motion over a wide range of temperatures, and the transduction becomes extremely nonlinear.
We numerically implement a model that describes transduction in this regime, in contrast to the conventional analytical description, which fails for fluctuations that approach the linewidth.
Moreover, we analyze how the nonlinear response of the radiation pressure force to stochastic fluctuations alters the shape of mechanical fluctuation spectra.
Finally, we provide a proof-of-concept demonstration of exploiting the nonlinearity to conduct sensitive quadratic readout of nanomechanical displacement.

\section{Results}
\subsection{Sliced photonic crystal nanobeam}
% 
% Figure 1
% 
\begin{figure*}[t!]
\includegraphics[width=\textwidth]{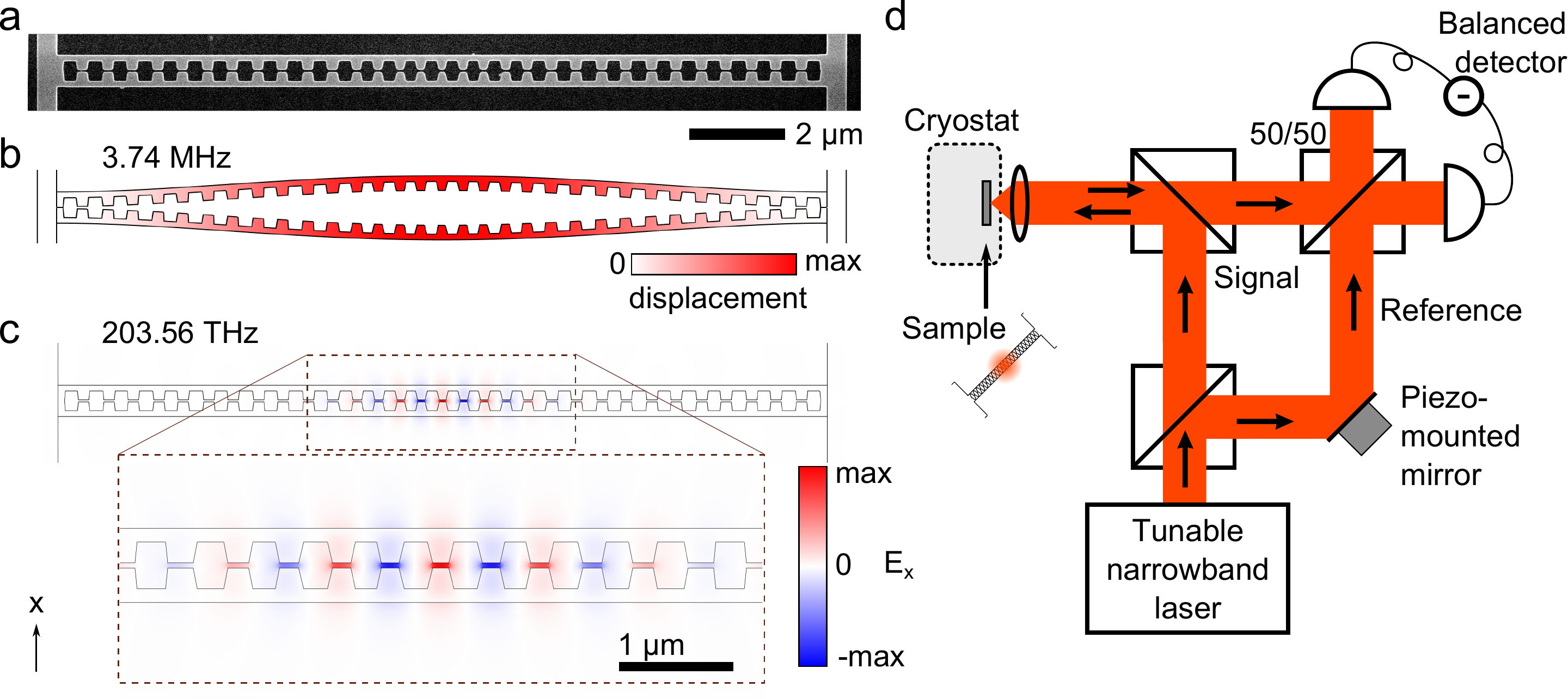}
\caption{{\bfseries Structure and setup.}
(a) Electron microscope image of a silicon sliced nanobeam.
The shown part is free-standing and has a thickness of 250 nm.
The scale bar is valid for panels (a--c).
(b) Simulated displacement profile of the fundamental mechanical resonance, which strongly modifies the gap size.
(c) Simulated transverse electric field of the fundamental optical cavity resonance.
The inset shows an enlarged view of the cavity region, formed by a tapered variation of the distances between, and sizes of, the holes.
(d) Schematic diagram of the employed balanced homodyne interferometer measurement setup.
The reflection from the sliced nanobeam is interfered with the light from the reference arm, enabling near-quantum-limited measurement of fluctuation spectra even with low power incident on the sample (see Methods for details).}
\label{setup_and_sample}
\end{figure*}

Figure~\ref{setup_and_sample}a shows the optomechanical system we employ.
It combines low-mass, MHz-frequency, nanomechanical modes with subwavelength optical field confinement in a sliced photonic crystal nanobeam\autocite{Leijssen2015}, to establish strong optomechanical interactions with photon-phonon coupling rates $g_0$ in the range of tens of MHz.
The fundamental mechanical resonance of the sliced nanobeam, shown in Fig.~\ref{setup_and_sample}b, strongly influences the gap distance $d$ in the middle of the beam. 
The motion of the resonator in this mechanical mode is associated with a simulated effective mass of 1.5~pg, leading to relatively large zero-point fluctuations $\xzpf = \sqrt{\hbar / 2 m \Om} = 43$~fm.
As shown in the optical field profile of the fundamental cavity resonance of the structure in Fig.~\ref{setup_and_sample}c, the nanoscale gap in the middle of the beam confines the light to a small area, which makes the optical cavity resonance frequency $\oc$ strongly dependent on the gap size $d$ between the two halves of the nanobeam.
With the fabricated gap size of 45--50 nm, we simulated the optical frequency change due to a displacement of the beams to be $\partial \omega / \partial x = 0.8$~THz/nm, where $x\equiv d / 2$.
This leads to an expected optomechanical coupling rate of $g_0/ 2 \pi = 35$~MHz.
We decrease the optical cavity decay rate and increase the outcoupling at normal incidence by engineering the angular radiation spectrum of the sliced nanobeam structure\autocite{Quan2011,Tran2010} (see Methods). 
The resultant simulated optical decay rate is 8.8~GHz, showing that the sliced nanobeam design is capable of combining large optomechanical interactions with low optical losses. % kappa_sim = 8.62-8.85 GHz;  Q_sim = 23100 -23600

We employ a balanced homodyne detection scheme, schematically shown in Fig.~\ref{setup_and_sample}d, to study the fluctuations imparted on the light in the nanobeam cavity through the optomechanical interaction (see Methods).
Figure~\ref{lw}a depicts fluctuation spectra recorded with an electronic spectrum analyser, showing the two fundamental mechanical resonances of one device, measured at 3~K with the laser on-resonance with the cavity.
We ascribe the two resonances to the two half-beams moving at slightly different natural frequencies.
The fact that in this device the two resonances are nearly of equal strength indicates that the two half-beams are mechanically coupled at a rate smaller than their intrinsic frequency difference\autocite{Leijssen2015}, and thus moving approximately independent of each other: stronger mechanical coupling would result in hybridization of the resonances into symmetric and antisymmetric eigenmodes with different optical transduction strengths.

%
% Linewidth broadening
%
\subsection{Modification of optical response}

%
% Figure 2
%
\begin{figure*}
  \includegraphics[width=\textwidth]{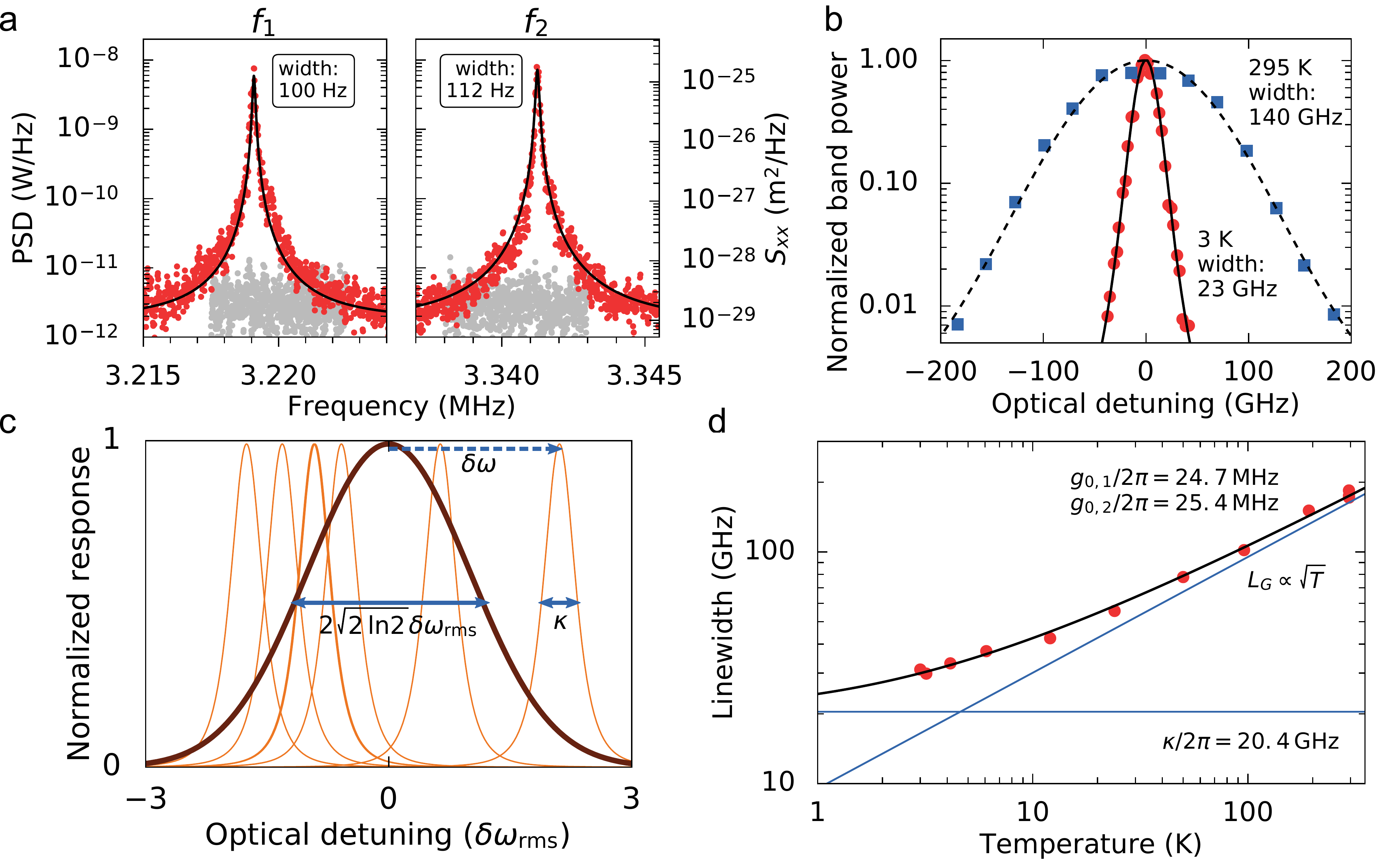}
\caption{{\bfseries Optical linewidth broadening.}
(a) Recorded optically measured spectra of the two fundamental mechanical resonances (PSD: power spectral density, optical power incident on sample: 11.3~nW).
Grey noise spectra were recorded with the signal arm of the interferometer blocked.
Black lines show the lorentzian fit, used to determine the linewidths (full width at half maximum) shown in the figure.
The displacement spectral density scale on the right-hand side assumes linear transduction of the known thermal motion of the structure at the cryostat temperature.
(b) Detuning dependence of the measured transduced thermal motion, measured as the band power at the fundamental mechanical frequency, at room temperature and at 3 K.
The black solid and dashed lines show fits with a Voigt lineshape squared (see Methods), with the widths (full width at half maximum) shown.
(c) Schematic representation of thermal motion-induced linewidth broadening: the thin orange lines represent the intrinsic cavity response at a few example detunings, while the thick brown line shows the overall response resulting from averaging over the fluctuating detuning.
(d) Optical linewidth versus temperature.
The solid line is a fit with a model that assumes a constant lorentzian intrinsic linewidth $\kappa$ convolved with a gaussian with a width $L_G$ that depends on $\sqrt{T}$ (details in text).
The asymptotes of the fit function allow us to extract $\kappa$ and the optomechanical coupling rate $g_0$, shown in the graph.
}
\label{lw}
\end{figure*}

The signal strength of the transduced motion around the cavity resonance wavelength of 1457.5~nm, measured at 3~K and at room temperature, is shown in Fig.~\ref{lw}b as a function of laser detuning.
The fluctuations are recorded while continuously sweeping the piezo-mounted mirror over multiple interferometer fringes, thus averaging the measured signal quadratures.
As we analyze in detail elsewhere\autocite{LaGala2016}, the resulting signal strength acquires a simple single-peaked detuning dependence with a maximum for the laser tuned to the cavity resonance (see also Appendix).
As Fig.~\ref{lw}b shows, the apparent linewidth of the optical resonance is strongly influenced by temperature.
We infer from this that the frequency fluctuations of the cavity due to the thermal motion of the mechanical resonator dominate the response, which occurs when they are larger than the intrinsic optical linewidth.
This is illustrated in Fig.~\ref{lw}c: While the intrinsic optical response of the cavity is Lorentzian with linewidth $\kappa$ (orange thin lines), the distribution of cavity frequency fluctuations due to thermal motion has a Gaussian spectrum (brown thick line), whose linewidth $L_G$ is related to the root-mean-square value of the frequency fluctuations $\dorms$, as $L_G =  2 \sqrt{2 \ln 2} \dorms$.
In the bad-cavity limit we consider here, we model the observed cavity response as a Voigt lineshape, which is a convolution between the Lorentzian cavity response and the Gaussian distribution of cavity resonance frequencies due to the Brownian motion.
We note that the measured electronic power spectral density is proportional to the square of the optical response, which leads to a smaller apparent linewidth in the detuning dependence shown in Fig.~\ref{lw}b.
In the following, we only report the extracted linewidth (see Methods), which directly corresponds to the optical loss rate $\kappa$ in the low-temperature limit, and the full width at half maximum of the frequency fluctuation distribution in the high-temperature limit.

As the thermomechanical displacement variance is given by $\varone{\xth^2}=2\nth\xzpf^2$, the induced frequency fluctuations due to a single mechanical mode at frequency $\Om$ are characterized by a root-mean-square amplitude
\begin{equation}
  \dorms \equiv \sqrt{\varone{\delta \omega^2}} = g_0 \sqrt{2 \nth } = g_0 \sqrt{2 k_B T / \hbar \Om},
  \label{omegarms}
\end{equation}
which reveals a square-root dependence on temperature.
In case multiple independent mechanical modes are coupled to the optical cavity, the variances of the cavity frequency fluctuations are added, i.e. $\dorms^2 = \sum_j \varone{\delta  \omega^2_j}$, which preserves the overall temperature dependence.
Fig.~\ref{lw}d shows the full measured temperature dependence of the apparent linewidth, which exhibits the expected square-root dependence on temperature at the higher temperatures.
We fit the datapoints using an equation that approximates the linewidth of the Voigt lineshape (see Methods), with a fixed Lorentzian contribution due to the intrinsic optical loss and a Gaussian contribution that follows Eq.~\ref{omegarms}.
The resulting fit curve is shown in Fig.~\ref{lw}d, together with its asymptotes (thin blue lines).
These asymptotes allow us to directly extract the intrinsic optical linewidth $\kappa$ and the variance of thermal motion-induced frequency fluctuations of the cavity, without further calibration.
There are two mechanical resonances that show significant coupling to the optical cavity resonance as shown in Fig.~\ref{lw}a, and we derive the ratio between their coupling strengths from the ratio between the signal strengths at the two mechanical resonance frequencies\autocite{Leijssen2015}.
Using this ratio and the measured resonance frequencies, we obtain $g_0/2\pi=24.7$~MHz and 25.4~MHz for the two mechanical modes. 

This unprecedented optomechanical coupling rate, combined with the extracted optical decay rate of $\kappa/2\pi=20.4$~GHz and the mechanical decay rate of $\Gamma/2\pi = 100$~Hz, means that this device has a single-photon cooperativity $C_0 = {4 g_0^2}/{\kappa \Gamma} = 1.1 \times 10^3$.
The single-photon cooperativity is a regularly used metric that combines optomechanical coupling and losses\autocite{Aspelmeyer2014,Schilling2016}.
It signals the inverse of the average number of intracavity photons that would be needed to perform a measurement at the standard quantum limit, if all photons escaping the cavity could be employed towards such a measurement.
Interestingly, the combination of quadrature-averaged detection with temperature-dependent linewidth measurement allows direct extraction of $C_0$, with no other calibration than that of the temperature of the mechanical bath.
The extremely high value we report here, which exceeds previously reported nano-optomechanical architectures by 2 to 3 orders of magnitude\autocite{Aspelmeyer2014,Meenehan2015,Schilling2016}, highlights the prospects of such systems for measurement-based quantum control of motion.

%
% Nonlinear transduction
%
\subsection{Nonlinear transduction}

%
% Figure 3
% 
\begin{figure*}
  \includegraphics[width=\textwidth]{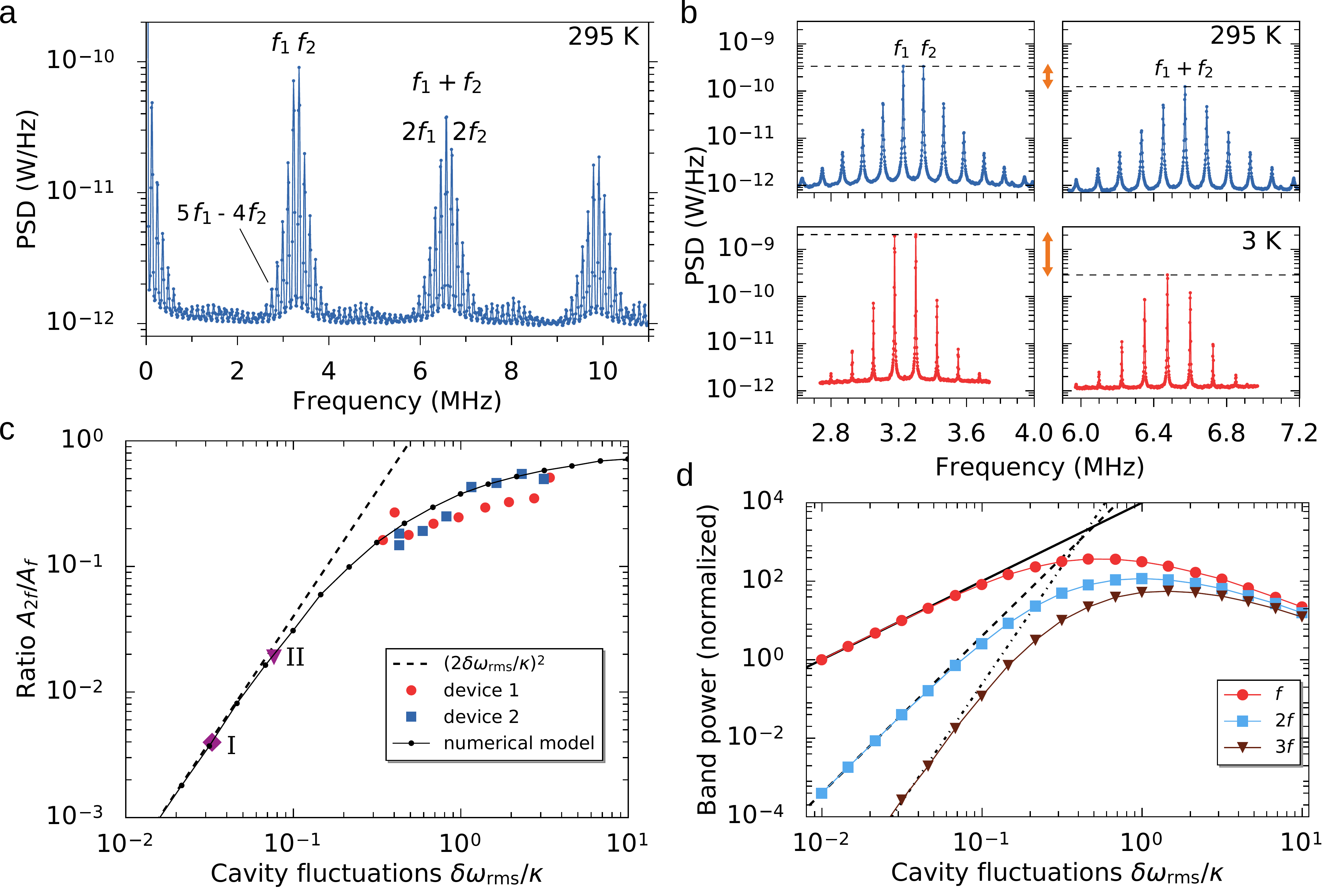}
\caption{{\bfseries Nonlinear transduction.}
(a) Power spectral density of transduced thermal motion, measured at room temperature (295 K).
(b) Power spectral density of the group of peaks around $f_1,f_2$ and around $f_1+f_2$, at room temperature (top) and at 3~K (bottom).
(c) Ratio between the 3 second-order peaks around $f_1+f_2$ and the peaks at the fundamental frequencies $f_1,f_2$ as a function of the relative cavity fluctuations $\delta \omega_\mathrm{rms} / \kappa$.
Device 1 is the same device presented in the other figures; device 2 is a similar device with different parameters (see text); room-temperature measurements from two recent publications are indicated with datapoints I (ref.~\cite{Brawley2016}) and II (ref.~\cite{Leijssen2015}).
The dashed line indicates the prediction from an analytical model with an order-by-order approximation, while the black datapoints connected with solid lines represent a calculation based on a numerically generated timetrace to simulate thermal motion.
(d) Expected band power at the fundamental frequency $f$ as well as at $2f$ and $3f$, as a function of the relative cavity fluctuations.
Solid, dashed, and dash-dotted lines follow the order-by-order approximation, while the colored points are the outcome of our numerical model.
}
\label{nonlinear}
\end{figure*}

When the relative cavity fluctuations $\dorms / \kappa$ are large, a typical mechanical oscillation samples the full width of its Lorentzian lineshape.
Since this lineshape is nonlinear, higher-order harmonics are expected to appear in the transduced optical fluctuation spectrum.
At room temperature, the cavity frequency fluctuations due to both mechanical modes have an amplitude $\dorms =\left( 2 \sum_j \overline{n}_{\mathrm{th},j} g_{0,j}^2 \right)^{1 / 2} = 2 \pi \times 69$~GHz $\approx 3.4 \kappa$.
%\todo{f1=3.22MHz,g0,1=2pi 24.7MHz; f2=3.34MHz,g0,2=2pi 25.4MHz; then $\dorms$ = 2 pi 68.6GHz = 3.36 kappa --- Number for only $f_2$: 2 pi 51 GHz = 2.5 kappa}.
As shown in Fig.~\ref{nonlinear}a, at room temperature the measurement signal indeed contains fluctuations at (mixed) integer multiples of the two fundamental mechanical resonances, i.e. $f_{j,k} =|j f_1 \pm k f_2 |$, where $j,k \in \{0,1,2,\ldots\}$.
Around the fundamental frequencies near $3.3$~MHz, we observe odd mixing terms up to 9th order (e.g. a peak is visible at $f_{5,-4} = 5 f_1 - 4 f_2$), and around the sum frequency at $6.6$~MHz, even mixing terms up to 10th order can be identified.

At a temperature of 3~K, the ratio between the cavity frequency fluctuations, caused by both mechanical modes, and the intrinsic optical linewidth is $\dorms / \kappa = 0.34$, which still leads to significant higher-order transduction.
Fig.~\ref{nonlinear}b shows a direct comparison of the spectra obtained at room temperature and at 3~K.
As a measure for the higher-order transduction, we take the ratio between the second- and first-order transduction.
We expect the ratio between the orders to be independent of other parameters, such that it gives direct insight in the strength of the nonlinearity.
This ratio is clearly larger at higher temperature (the inverse of the ratio is indicated with arrows in Fig.~\ref{nonlinear}b).

Higher-order transduction has previously been described with an analytical model that is based on a Taylor expansion of the measurement output around the average detuning\autocite{Doolin2014,Leijssen2015,Brawley2016}.
For small frequency modulation, the higher-order terms in this expansion can be approximated as independent.
Mathematically, this is based on an order-by-order approximation $(\cos \Om t)^n \approx 2^{-(n+1)}\cos n \Om t$.
The resulting expression for the maximum signal power measured at the (multiple of the) resonance frequency $k \Omega$ is
\begin{equation}
  \varone{P^2}_{k \Omega} = 2 A^2 k! \left( \frac{2 \varone{\delta \omega^2}}{\kappa^2} \right)^k,
  \label{eq:orderbyorder}
\end{equation}
where $A$ is a constant factor that depends on the optical power used as well as the coupling efficiency to the cavity (for details, see Appendix and ref.~\cite{LaGala2016}).

In Fig.~\ref{nonlinear}c, the dashed line shows the ratio between the second- and first-order transduction that follows from this order-by-order approximation, $(2 \dorms / \kappa)^2$.
The datapoints labeled I and II, which represent room-temperature measurements reported in refs.~\cite{Brawley2016} and \cite{Leijssen2015}, respectively, are still well-explained by the order-by-order approximation.
For the devices under study, $\dorms / \kappa$ is so large that this approach breaks down.
This is shown by our temperature sweep data, where we take the fitted area $A_{f_j}$ under the peaks at the fundamental frequency and at twice the fundamental frequency (labeled in Fig.~\ref{nonlinear}a), and plot their ratio: $(A_{2f_1} + A_{f_1+f_2} + A_{2f_2}) / (A_{f_1} + A_{f_2})$. 
The red circles indicate the results for the same device presented in the other figures, while the blue squares represent a different device with $g_0 / 2 \pi = 10.8$~MHz and $\kappa / 2 \pi = 9.0$~GHz.

Where the order-by-order approximation breaks down, the nonlinear transduction still follows the prediction of our full model, in which we calculate the expected homodyne output from a numerical time-domain simulation of the thermal motion (see Methods).
We then Fourier transform the simulated signal and plot the ratio of the band powers, noting that the numerical result does not depend on whether we include one or two mechanical resonances when plotted against total relative cavity fluctuations $\dorms/ \kappa$.
This full model prediction is shown with the black dots connected by solid lines in Fig.~\ref{nonlinear}c.
The numerical simulation follows the experimentally observed trend, and shows that the ratio between second- and first-order transduction saturates close to unity at higher fluctuation amplitudes.
It is important to note that only such a numerical approach correctly takes into account the statistics of the transduced motion: the strongly nonlinear conversion of displacement to optical field precludes an analysis based solely on one or several moments of the displacement distribution.
In other words, a single peak in the fluctuation spectrum at $k \Om$ can no longer be dominantly associated with a specific scattering process involving $k$ phonons, but also contains contributions due to $k+2,k+4,\ldots$ phonons.

Using the numerical model, we calculate the absolute signal power due to thermal motion at the fundamental frequency $f$ and its multiples $2f$ and $3f$ (points in Fig.~\ref{nonlinear}d). 

For cavity frequency fluctuations $\delta \omega_\mathrm{rms} / \kappa$ larger than about 10\%, both linear and higher-order transduction no longer follow the order-by-order approximation given in equation~\ref{eq:orderbyorder} (black lines), which predicts monotonic increases in signal strength.
Instead, we recognize an optimum single-harmonic signal strength due to cavity frequency fluctuations, beyond which larger fluctuations cause the cavity to be off-resonance most of the time.
This reduces the transduction at a single harmonic in the optical signal, instead distributing energy equally among an increasing number of harmonics as the system operates deeper in the nonlinear regime.

\subsection{Quadratic measurement of motion}

%
% Figure 4
%
\begin{figure}[t!]
\includegraphics[width=0.50\textwidth]{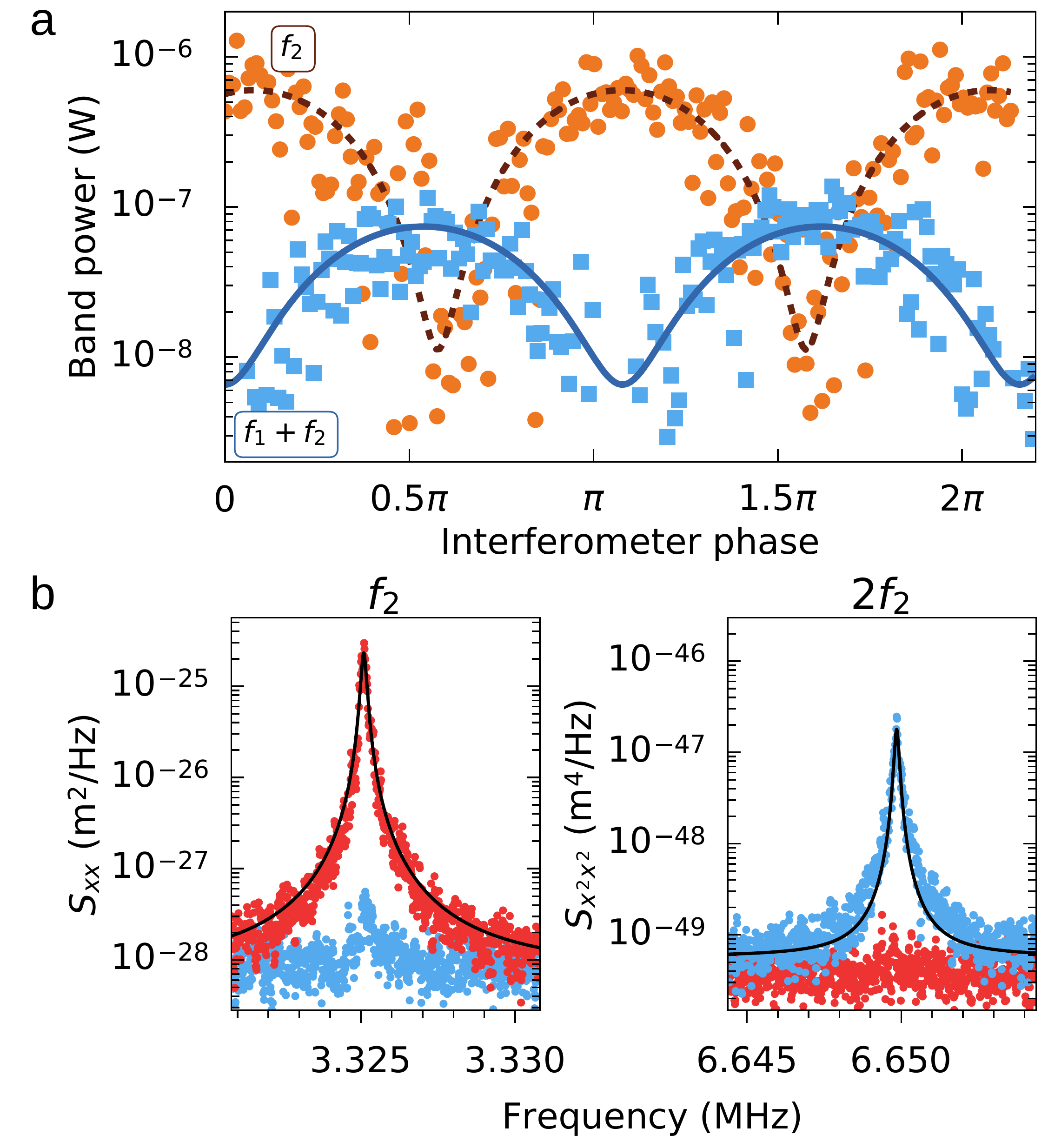}
\caption{{\bfseries Quadratic measurement.}
(a) Transduction of first and second order peaks (orange and lightblue datapoints, respectively) while sweeping the piezo mirror position, with 10.5 nW optical power incident on the sample.
The horizontal scale was derived from the sinusoidal fit to the band power of $f_2$ shown in the figure, and the data at $f_1+f_2$ was shifted horizontally by about 3\% to compensate for measurement drift.
(b) Spectra at $f_2$ and $2f_2$ at optimal mirror positions for linear and quadratic transduction (red and lightblue datapoints, respectively), taken with 12.8 nW of incident optical power.
The black solid lines are Lorentzian fits, whose area was used to calibrate the vertical scale.
}
\label{firstordersecondorder}
\end{figure}

A measurement that is directly sensitive to the square of displacement, $x^2$, is proportional to the energy of the mechanical resonator.
Displacement-squared optomechanical coupling hence provides a means to perform quantum non-demolition measurements of phonon number\autocite{Thompson2008,Miao2009,Clerk2010,Paraiso2015,Kaviani2015}.
With sufficiently suppressed linear backaction, measurements of $x^2$ have been proposed as a possible route to preparing non-classical states of motion of the mechanical resonator, conditional on the outcome of such a nonlinear measurement\autocite{Vanner2011d,Doolin2014,Brawley2016}.
Nonlinear transduction allows measurements of $x^2$ by detecting the transduced motion at twice the fundamental frequency, $2f$.
This gives rise to an effective quadratic coupling rate given by $g_0^2 / \kappa$ (ref.~\cite{Vanner2011d}), which amounts to $2 \pi \times 32$~kHz in the device presented here.
To put that number in perspective, one can compare it to the quadratic coupling rate $\mu_0 = (\partial^2 \oc / \partial x^2) \xzpf^2$ in devices in which the frequency is not linearly dependent on the displacement. 
A state-of-the-art double-slotted photonic crystal system recently demonstrated\autocite{Paraiso2015} a quadratic coupling rate of $\mu_0 / 2 \pi = 245$~Hz.

In Fig.~\ref{firstordersecondorder} we exploit the large nonlinear transduction in our device to show selective linear and quadratic measurements of mechanical displacement at 3~K.
Figure~\ref{firstordersecondorder}a shows the strength of the first- and second-order transduction (proportional to $x$ and $x^2$, respectively) as a function of the piezo mirror position that controls the homodyne phase, which is now no longer continuously swept.
The data follows the expected sinusoidal dependence, as evidenced by the fits shown with solid lines.

We obtained the spectra in Fig.~\ref{firstordersecondorder}b by positioning the piezo mirror at the optimum points for first- and second-order transduction, depicted by the red and blue datapoints respectively.
The vertical scale was calibrated by using the order-by-order approximation, where the area under the peaks in the spectrum is proportional to the variance of the thermal motion of the structure, which here provides a lower limit for the sensitivity. 
This analysis yields an imprecision for the displacement-squared measurement of $2.0 \times 10^{-25}$~m$^2$/Hz$^{1 / 2}$.
% Brawley / Vanner quoted 3.3e-24 m2 / sqrt(Hz)
% 4.0e-50 m4 / Hz 
% corresponds to imprecision of $119 \xzpf^2$/Hz$^{1 / 2}$
% compared to Brawley, Vanner et al: imprecision in m2 / sqrt(Hz) is 16.5 times better, in xzpf2 / sqrt(Hz) it's 500 times smaller, the phonon detection limit is 47 times better
With the current mechanical linewidth and this level of imprecision we would be able to measure a phonon occupation of 752 with a signal-to-noise ratio of 1, which is an improvement over the measurement sensitivity in the proof-of-concept experiment reported in ref.~\cite{Brawley2016} by approximately a factor 50.
In the data shown here, the suppression of linear transduction is 28~dB, which could be further improved by implementing a feedback loop to lock the homodyne interferometer phase to a desired value.

It is important to note that our numerical model, as shown in Fig.~\ref{nonlinear}d, predicts second-order transduction at only 10\% of the strength of the order-by-order approximation for $\dorms / \kappa \approx 0.34$, which is the size of the relative cavity fluctuations in our device at 3~K.
Therefore, we expect that if this device is cooled further, to the point where the order-by-order approximation accurately predicts the transduction, an even lower imprecision noise of $6.3 \times 10^{-26}$~m$^2$/Hz$^{1 / 2}$ would be recovered without any further optimization.
% imprecision: 4.0e-51 m4 / Hz or 1.41e3 $\xzpf^4 / \mathrm{Hz}$ or 37.6 $\xzpf^2 / \mathrm{Hz}^{1 / 2}$
This would correspond to measuring a phonon occupation of 238 with unity signal-to-noise.
In the regime where the order-by-order approximation holds, this phonon occupation detection limit for quadratic measurement scales as
\begin{equation}
  \frac{1}{\overline{n}_\mathrm{min}} = 16 \left( \frac{g_0}{\kappa} \right)^2 \eta \sqrt{\frac{\Pin / \hbar \oc}{\Gamma}},
\end{equation}
with $\eta$ the coupling efficiency to our cavity, and $\Pin$ the incident optical power (see Appendix).
This allows estimating what improvements we need to reach $\overline{n}_\mathrm{min} \approx 1$, or a sensitivity that enables performing a measurement of phonon number near the quantum limit.
This would for example be reached by improvements in $g_0$ and $\kappa$ by factors of $\sqrt{2}$ and $2$, respectively, as well as increasing the light collection efficiency $\eta$ to 40\% from the 1.3\% estimated in the current device (see Appendix), by employing waveguide-based coupling strategies\autocite{Groblacher2013}.
This shows that with modest improvements of the device design and changes in the detection method, achieving quadratic measurement with an imprecision near the quadratic zero-point fluctuations is within reach.

\subsection{Radiation pressure force with large cavity frequency fluctuations}
%
% Figure 5
%
\begin{figure*}
  \includegraphics[width=\textwidth]{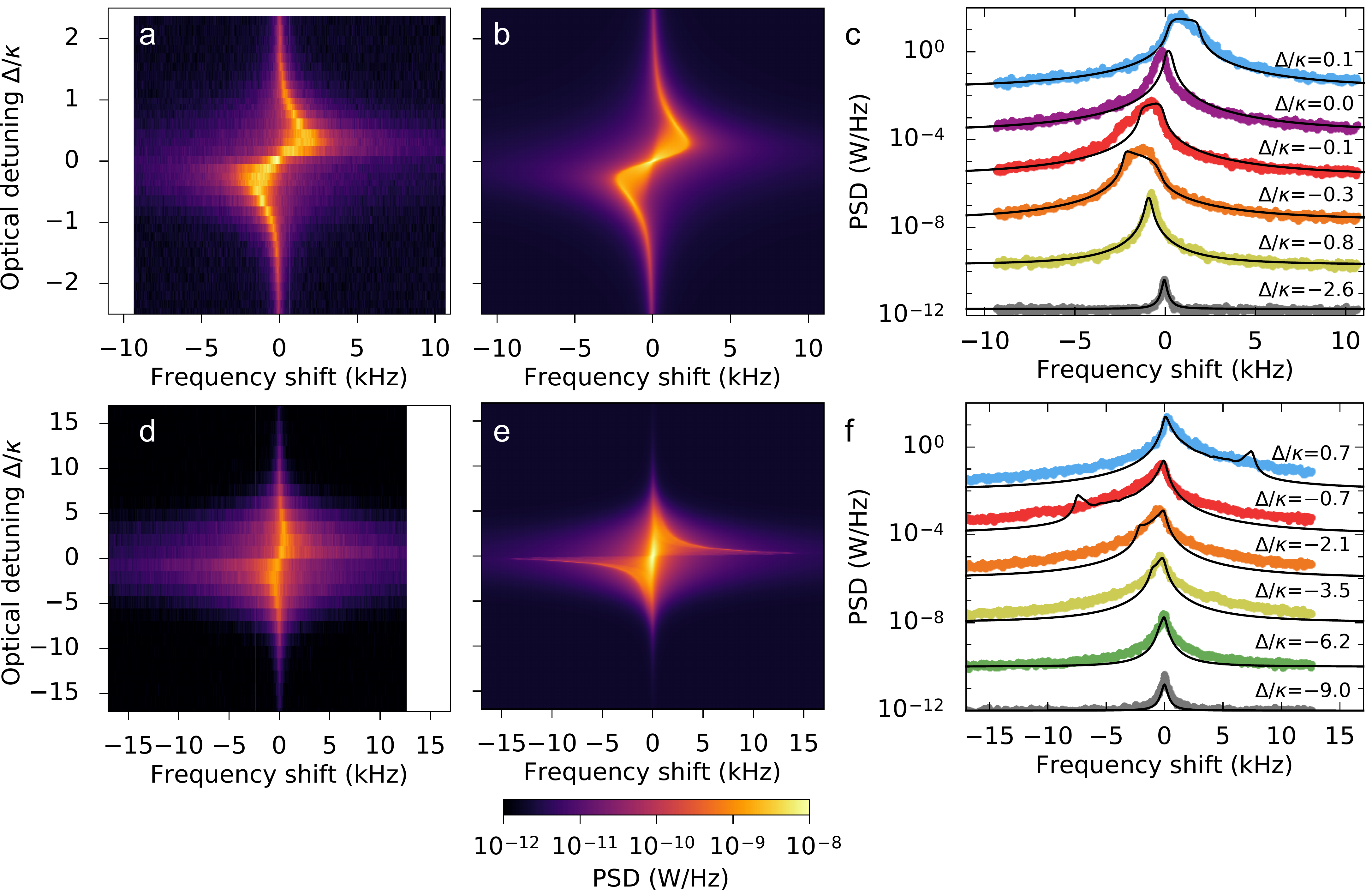}
\caption{{\bfseries Optical spring effect modified by large cavity frequency fluctuations.}
(a,d) Experimental spectrograms, showing transduced mechanical spectra (horizontal axis) versus the optical laser frequency (vertical axis), at two different experimental conditions: (a) cooled at 3 K, incident laser power 20.6~nW; (d) room temperature, 295 K, with incident laser power 124~nW.
(b,e) Simulated spectrograms corresponding to those shown in (a,d), obtained by calculating frequency shifts for a large number of motional amplitudes sampled from a thermal distribution, and averaging Lorentzian lineshapes with center frequencies given by the calculated shifts (see text for details).
The colorbar below panel (e) is shared for all the spectrograms (PSD: power spectral density).
(c,f) Individual spectra at various detunings, overlaid with the corresponding simulated spectrum.
Spectra were offset by a factor of $10^2$ between them to avoid overlap.
}
\label{optspring}
\end{figure*}

The nonlinear regime not only impacts optical transduction of motion, but also pronouncedly affects the mechanical fluctuations themselves through its influence on radiation pressure backaction.
The strongest manifestation of backaction in the regime where the cavity reacts nearly instantaneously to the mechanical motion ($\kappa \gg \Om$) is the \emph{optical spring effect}, which alters the mechanical resonance frequency depending on the detuning between a drive laser frequency and the cavity resonance\autocite{Aspelmeyer2014}.
Figure~\ref{optspring}a and d show experimentally obtained spectrograms at low temperature (3 K, incident optical power 20.6~nW) and at room temperature (295 K, incident optical power 124 nW) around the fundamental harmonic of one of the mechanical modes, while Fig.~\ref{optspring}c and f show several crosscuts at selected detunings.
At low temperature, the observed effect is very similar to the normal (linearized) optical spring effect, where a blue detuned laser shifts the mechanical frequency upwards and vice versa for red detuning.
However, we additionally observe a broadening of the obtained spectra both when the laser is red- and blue-detuned by $\sim$10~GHz ($\kappa/2$).
At room temperature, the spectra also become asymmetric: instead of a symmetric Lorentzian lineshape, the peak in spectra close to the optical resonance has a much steeper edge on one side than on the other side.
In addition, the shift of the peak due to the optical spring effect scales linearly with the incident optical power, but the dataset at room temperature shows a smaller, rather than a larger shift than the dataset at 3~K, even though the used optical power is 5 times higher. 

To account for these observations, we again need to take the large fluctuations of the cavity frequency due to thermal motion into consideration.
The typical optomechanical model for the optical spring effect is based on the linearized equations of motion\autocite{Aspelmeyer2014}, which don't apply in the nonlinear regime we reach here.
Instead, we calculate an effective spring constant from the first fourier coefficent of the radiation pressure force, while the resonator oscillates harmonically, $x(t) = x_0 \cos \Omega t$.
For this, we express the radiation pressure force in the cavity as 
\begin{equation}
F_\mathrm{rad} = \frac{\hbar ( \partial \oc / \partial x ) n_c^\mathrm{max}}{1+u^2},
\end{equation}
where $u \equiv\frac{2}{\kappa} (\overline{\Delta} + (\partial \oc / \partial x) x)$ with $\overline{\Delta}$ the average detuning between the laser and cavity frequency, and $n_c^\mathrm{max}$ is the maximum number of photons in the cavity when it is driven at resonance.
The effective spring constant can be directly rescaled to obtain the frequency shift (see Appendix).
As before, we numerically sample a thermal distribution to model the behaviour of the system, in this case by averaging over a large number of different amplitudes $x_0$.
Finally, we average a large number of Lorentzian lineshapes with the center frequencies given by the simulated frequency shifts and their widths set by the median linewidth measured in the experiment.
The resulting simulated spectra are shown in Fig.~\ref{optspring}b,e and plotted as black solid lines in Fig.~\ref{optspring}c,f.
To allow overplotting the simulated spectra with the experimental crosscuts, we rescaled all the simulated spectra such that the maximum value of the spectrum closest to resonance matched the experimental value.
The model reproduces the broadening of the peaks as well as the asymmetry for the high-temperature data.
At room temperature, a strong edge at larger frequency shifts is observed in the model, which is absent in the measurements.
We attribute this to the presence of the other mechanical mode, which will tend to soften this effect.
An interesting feature in this model is that the spectra obtained with the laser on-resonance are not affected at all, and we indeed observed experimentally that at $\overline{\Delta} / \kappa=0$ the effect is much smaller if not entirely absent.

\section{Discussion}

Our analysis of both nonlinear transduction and backaction relies on a numerical model instead of an analytical description, as the typical order-by-order approximation to describe transduction breaks down at large relative cavity fluctuation strengths.
This breakdown is related to the fact that the Taylor expansion used to describe the intracavity field\autocite{Brawley2016} as a function of the relative detuning $u \equiv\frac{2}{\kappa} (\overline{\Delta} + \delta \omega)$,
\begin{equation}
  a %= \frac{\sqrt{\eta \kappa}}{i \Delta + \kappa / 2}
  %= \frac{\sqrt{n_c^\mathrm{max}}}{1+2 i \delta \omega / \kappa}
  = \frac{\sqrt{n_c^\mathrm{max}}}{1+ i u}
  \approx \sqrt{n_c^\mathrm{max}} (1 - i u - u^2 + i u^3 + u^4 \ldots),
  \label{eq:expansion}
\end{equation}
does not converge for $| u | > 1$. For example, at average detuning $\overline{\Delta} = 0$, $u = 2 \delta \omega / \kappa$ represents the cavity frequency fluctuations due to the thermal motion, meaning the power series does not converge when the mechanical motion changes the cavity frequency by more than a linewidth.
Therefore, this model cannot be used to extrapolate the expected sensitivity of optomechanical measurements in the nonlinear regime we describe here.
Our numerical model relies on a direct calculation of the optical response to mechanical motion, which can be performed for any cavity frequency change $\delta \omega$.
It is however crucial to correctly simulate the distribution of mechanical amplitudes within the thermal mechanical state.

We have demonstrated the effects of the optomechanical nonlinearity in a system that operates in the regime $\sqrt{2 \nth} g_0/ \kappa \gtrsim 1$, and described how it modifies the optical response, transduction, and backaction over a wide range of temperatures.
It is to be expected that a growing number of optomechanical systems will operate in this regime as parameters continue to improve, and the single-photon strong coupling regime ($g_0 > \kappa,\Gamma$) is approached.
Indeed, various characteristics of the nonlinear regime we demonstrate are reminiscent of the expected effects due to quantum fluctuations in the single-photon strong coupling regime, including a modification of the optical response and the appearance of strong higher-order sidebands in optical fluctuation spectra. 
Whereas our sliced photonic crystal nanobeam devices demonstrate these effects for the bad-cavity limit, the regime has equally important impact for devices in the resolved-sideband regime ($\kappa \ll \Om$), although precise manifestations are expected to vary.
In particular, the optical excitation spectrum will be altered by displaying multiple discrete sidebands, as analysed in the strong-coupling regime for ground-state motion in ref.~\cite{Rabl2011}.
Moreover, backaction forces acquire an additional delay factor due to the longer lifetime of cavity excitation, which will lead to fluctuating damping and driving forces affecting the motion.
Notably, a combination of such dynamical backaction and the optomechanical nonlinearity could be used as a resource to manipulate thermal fluctuations beyond a Gaussian distribution.
Similar aims in the bad-cavity limit could be reached through nonlinear measurement, such as the displacement-squared measurements we demonstrated. 

Interestingly, the alteration of the optical response due to thermal motion provides a new method to directly determine the intrinsic optical linewidth $\kappa$ and the optomechanical coupling rate $g_0$: the ratio of different harmonics of the transduced mechanical spectrum allows retrieving the relative frequency fluctuations $\delta\omega_\mathrm{rms}/\kappa$.
Together with a measurement of the optical excitation linewidth (see Methods eq.~\ref{eq:voigtlw}) this uniquely determines both $g_0$ and $\kappa$.
This method only requires further knowledge of the bath temperature and the mechanical frequency. We note that there is no need for any other calibration, including characterization of optical powers.

The large single-photon cooperativity $C_0$ in the structures we present here offers prospects beyond the exploitation of the optomechanical nonlinearity, in particular for quantum measurement and control of mechanical motion.
For example, the requirement for feedback cooling to the ground state (thermal occupancy below 1) is that the cooperativity $C_0 n_c > \nth/(9\eta-1)$, where $n_c$ is the number of photons in the cavity\autocite{Genes2008,Krause2015a,Wilson2015}. 
In our sliced nanobeam structure, the measurement sensitivity is currently limited by the collection efficiency, which we estimate to be $\eta\approx1.3$\% (see Appendix).
The double-period modulation method we use to improve the coupling efficiency of a photonic crystal nanobeam cavity at normal incidence has previously been shown\autocite{Tran2010} to yield a simulated collection efficiency of more than 20\%.
Therefore we expect that by either improving the optical design, or by employing waveguide-based coupling schemes\autocite{Groblacher2013,Krause2015a}, the coupling can be further increased in our current free-space set-up.
For example, with a detection efficiency of 2/9, ground state preparation would be in reach at a very modest minimum cavity occupation of $n_c>17$ photons with the demonstrated parameters.
Finally, the large optomechanical coupling strength in combination with low loss provides other opportunities for measurement-based control in the bad-cavity limit, such as conditional state preparation of the mechanical resonator by pulsed measurements\cite{Vanner2011}, or quantum state swapping between the optical and mechanical degrees of freedom\cite{Bennett2016}.

\section*{Methods}

\subsection*{Sliced photonic crystal nanobeam design}
Eigenfrequency calculations were performed using finite element software (COMSOL Multiphysics).
For optical simulations, perfectly matched layers were used to allow the extraction of the optical radiation losses.
To estimate $\partial \oc / \partial x$, the optical frequency shift due to mechanical motion, $x$ (half of the gap size) was increased by 1 nm over the full length of the nanobeam, leading to a change in the simulated eigenfrequency.

The sliced nanobeam design confines light in the transverse direction due to total internal reflection, while along the nanobeam the photonic crystal patterning creates a bandgap for light.
An optical cavity is formed by a defect region in the middle with holes of different shape and periodicity, which is tapered to the periodic outer region over 5 holes to minimize optical losses\autocite{Quan2011}.
Additionally, we create a low-efficiency outcoupling grating in the structure by making the hole sizes alternatingly 5\% wider and narrower.
This double-period modulation allows part of the cavity field to scatter out at normal incidence\autocite{Tran2010}, which is efficiently collected by our free-space optical measurement setup.
Our experimental results indicate that this strategy increases the coupling efficiency to the optical cavity from approximately 0.1\% to 1--2\% (see Appendix).

\subsection*{Fabrication}
Devices were fabricated from a silicon-on-insulator substrate (SOITEC), with a 250~nm silicon device layer on top of 3~\textmu m silicon oxide.
Patterns were written using e-beam lithography in a 80~nm layer of spincoated HSQ resist (FOX-15, Dow Corning) and developed using TMAH.
The silicon layer was etched in an ICP plasma etcher using a combination of Cl$_2$ and HBr/O$_2$ gases.
Finally, the nanobeams were released to be free-standing by wet etching with HF, which also dissolves leftover resist and oxide-based deposits formed during plasma etching.
The wet etch was followed by critical point drying to prevent collapse of the nanobeams.
The device layer of SOI wafers typically contains compressive stress, which can induce buckling of the nanobeams even when using critical point drying.
We avoided this by incorporating stress-relief features in the support structure around the free-standing nanobeams.

\subsection*{Balanced homodyne detection at cryogenic temperatures}
A closed-cycle cryostat (Montana Cryostation C2) was used to control the sample temperature between 3 and 300~K.
%Using two 0.2 mm thin glass windows, the sample was positioned to allow free-space optical access while being shielded from thermal radiation.
We used an aspheric lens positioned outside the cryostat window with an effective focal length of 8 mm and a numerical aperture of 0.55 to focus the laser beam (New Focus Velocity 6725, linewidth $\leq$ 200~kHz) on the sample and to collect the reflection in free space.
A balanced detector with two nominally identical photodiodes (New Focus 1817-FS) detected the output of the homodyne interferometer, schematically shown in Fig.~\ref{setup_and_sample}c.
The detector signal was then Fourier transformed and the spectrum recorded with an electronic real-time spectrum analyser (Agilent MXA).
The optical power in the reference arm was 135~\textmu W or more, ensuring that the optical shot noise was at least as large as the electronic noise.
The pressure in the cryostat was typically 0.3~mbar at room temperature, and well below $10^{-4}$~mbar at cryogenic temperatures.

A measurement of the power dependence of the transduced signal (see Appendix) was performed to verify that the measurement of the thermal motion of the nanobeam is not influenced by additional heating by the laser beam.
We then used the temperature sensor placed next to the sample in the cryostat to calibrate the scale for the displacement power spectral density $S_{xx}$ in Fig.~\ref{lw}a and Fig.~\ref{firstordersecondorder}b.

\subsection*{Power spectral density and Voigt linewidth}
We plot the power spectral density of the electronic output signal of our measurement setup, which has units of W/Hz.
This corresponds to a power spectral density of a (virtual) optical power $P$, $S_{PP}$, which has units of W$^2$/Hz.
As a consequence, the Lorentzian detuning dependence of the optomechanical transduction leads to a Lorentzian-squared dependence in the power spectral density.
Similarly, we observe the square of the Gaussian distribution of cavity frequencies due to thermal motion.
Therefore, we fit the square of a Voigt lineshape, which models the convolution of a Lorentzian and a Gaussian lineshape, to our data, and extract the linewidth of the non-squared Voigt lineshape.
This value then directly corresponds to either the optical loss rate $\kappa$ or the amplitude of the frequency fluctuations $\dorms$, in the respective limits where $\kappa \ll \dorms$ or vice versa.

We use an empirical equation for the linewidth of a Voigt lineshape\autocite{Olivero1977}:
\begin{equation}
  0.5346 \kappa + \sqrt{0.2166 \kappa^2 + 8 \ln{2} \dorms^2},
  \label{eq:voigtlw}
\end{equation}
where $\kappa$ is the linewidth of the Lorentzian lineshape and $2 \sqrt{2 \ln 2}\dorms$ is the linewidth of the Gaussian lineshape.
To fit the linewidth as a function of temperature, we substitute $\dorms = \sqrt{2 g_0^2 k T / \hbar \Om}$.
To account for multiple independent mechanical modes, we can calculate the individual optomechanical coupling rates from the asymptote of the fit curve if we know the ratio between the transduced peaks in the spectrum\autocite{Leijssen2015}.
The variances due to the modes $j$ add up, $\dorms^2 = \sum_j \varone{\delta \omega^2_j}$, which means Eq.~\ref{omegarms} is modified to 
\begin{equation}
  \dorms^2 = \sum_j \left( \frac{g_{0,j}^2}{\hbar \Omega_{\mathrm{m},j}} \right) 2 k_B T.
\end{equation}

\subsection*{Numerical model for higher-order transduction}
Timetraces for the mechanical displacement $x$ were generated for one or two resonance frequencies $\Om$.
To simulate thermal motion, we randomly change the amplitude $A$ and phase $\varphi$ of harmonic motion $x=A \cos (\Om t + \varphi)$.
The points at which $A$ and $\varphi$ are changed are taken from a Poissonian distribution with mean time between jumps taken to be the mechanical damping time $\Gamma$.
The new amplitude is taken from an exponential distribution characterized by a mean proportional to the average thermal occupation $\nth$, while the phase is taken from a uniform distribution.
A timetrace of the measurement output $P$ was generated from the position timetrace using our full model for the transduction in the phase-averaged homodyne measurement, as given in the Appendix.
The discrete Fourier transform of this timetrace allowed us to extract the signal strength at $\Om$ and at higher-order multiples or mixing terms.

\printbibliography
\end{refsection}

\section*{Acknowledgements}
We thank Thijs Kleijntjens for assistance in setting up the balanced homodyne interferometer.
This work is part of the research programme of the Netherlands Organisation for Scientific Research (NWO).
E.V. gratefully acknowledges an NWO-Vidi grant for financial support. 
J.T.M. thankfully acknowledges funding from the European Union's Horizon 2020 research and innovation programme under the Marie Sklodowska-Curie grant agreement No 707364.

\clearpage

\section*{Appendix}
\begin{refsection}
\setcounter{figure}{0}
\renewcommand{\thefigure}{S\arabic{figure}}
\setcounter{equation}{0}
\renewcommand{\theequation}{S\arabic{equation}}

\subsection*{Balanced homodyne interferometric measurement of an optomechanical cavity}
The optical response of the balanced homodyne interferometer probing our cavity optomechanical system is a function of the average detuning $\overline{\Delta}$ between the laser frequency and the cavity frequency, the frequency shift of the cavity due to mechanical motion $\delta \omega$, the homodyne phase $\theta$, the cavity linewidth $\kappa$, and various constant factors such as the coupling efficiency to the cavity, and the optical power used for the measurement.
To obtain the full expression, we start by describing the reflection from our nanobeam cavity using input-output theory:
\begin{equation}
s_c = s_\mathrm{in} \left( c e^{i \phi} - \frac{\eta \kappa}{i \Delta + \kappa / 2} \right),
\end{equation}
with $\eta$ the coupling efficiency.
The term $c e^{i \phi}$ is due to non-resonant scattering from the nanobeam structure or the substrate.
If $\phi$ is 0, the resulting reflectivity shows a Lorentzian response, while other values will lead to the more general case of a Fano lineshape.

Using balanced homodyne detection, with the light in the reference arm, or local oscillator, described as $s_\mathrm{LO} = |s_\mathrm{LO}| e^{i \theta}$ and the input to the cavity $s_\mathrm{in}$ taken to be real, the output of the detector is proportional to a virtual optical power $P$:
\begin{equation}
\begin{aligned}
\frac{P}{\hbar \omega}&= \left| \frac{i}{\sqrt{2}}s_c +\frac{1}{\sqrt{2}}s_\mathrm{LO}\right|^2 - \left| \frac{i}{\sqrt{2}}s_\mathrm{LO}+\frac{1}{\sqrt{2}}s_c\right|^2\\
&= i(s_\mathrm{LO}s_c^*-s_\mathrm{LO}^*s_c)\\
&= |s_\mathrm{LO}| |s_\mathrm{in}| \left( -2 c \sin (\theta - \phi) + \frac{\eta \kappa \left(2 \Delta \cos \theta + \kappa \sin \theta \right)}{\Delta^2 + \kappa^2/4} \right).
\end{aligned}
\label{eq:homodyne}
\end{equation}
Finally, we can substitute $\Delta \equiv \overline{\Delta} + \delta \omega = \overline{\Delta} + \frac{\partial \omega}{\partial x} x$ to obtain the relationship between the measurement output $P$ and the displacement $x$.

\subsection*{Optomechanical transduction with order-by-order approximation}
We consider the Taylor expansion of the measurement output $P$ for small fluctuations $\delta \omega$ around $\overline{\Delta}$:
\begin{equation}
  P(\overline{\Delta}+\delta \omega) = P(\overline{\Delta}) + \sum_{k=1}^{\infty} \frac{\delta \omega^k}{k!} \frac{\partial^k P}{\partial \omega^k}.
\end{equation}
For harmonic fluctuations $\delta \omega = \delta \omega_0 \cos \Om t$, the individual terms of this expansion contribute at different frequencies.
To leading order, $\cos^k \varphi = \frac{1}{2^{k-1}} \cos k \varphi$, which means the higher-orders are completely spectrally separated, and we can consider this expansion \emph{order by order}.

We measure the power spectral density $S_{PP} (\Omega)$, whose integral over frequency gives the variance $\varone{P^2}$.
%, at the cost of up to 3 dB reduction of the theoretical maximum sensitivity.
%This effectively averages over the different quadratures of the light field reflected from the sample\todo{simpler: averaging over different phases of the interferometer}.
%While this decreases the maximum sensitivity to a cavity frequency modulation by up to 3 dB from the theoretical sensitivity at optimal detection quadrature, it removes all dependence of the tranduction strength on the shape of the reflectivity spectrum, which is governed by non-resonant scattering in the reflected light whose phase and amplitude depend unpredictably on sample and experimental alignment.
%In particular, the resulting signal strength using quadrature averaging acquires a simple single-peaked detuning dependence with a maximum for the laser tuned to cavity resonance, as we analyze in detail elsewhere\autocite{LaGala2016}.\todo{this paragraph reads purely as methods.Could we introduce already a result in it?}
Averaging over the homodyne phase, the detuning dependence for $(\partial^k P / \partial \omega^k)^2$ has a simple Lorentzian lineshape, raised to the power $(k+1)$, as we analyze elsewhere\autocite{LaGala2016}.
The maximum contribution is therefore at resonance ($\overline{\Delta} = 0$) and can be expressed as
\begin{equation}
  \left( \frac{\partial^k P}{\partial \omega^k} \right)^2 = A^2 k!^2 \left( \frac{2}{\kappa} \right)^{2k},
\end{equation}
where $A^2 = 8 \Pin \Plo \eta^2$ is a constant prefactor that depends on the optical powers $\Pin,\Plo$ in the signal and reference arm, respectively, and on the coupling efficiency to the cavity $\eta$.

We now calculate the band power at a frequency $k \Om$, where only the $k$th term of the Taylor expansion contributes, as
\begin{equation}
  \varone{P^2}_{k \Om} \equiv \int_{k \Om} S_{PP} d\Omega = A^2 \left( \frac{2}{\kappa} \right)^{2k} \varone{\left(\delta \omega^k \right)^2}_{k \Om}
  \label{eq:varPfromvarOm}
\end{equation}
Since $\delta \omega = (\partial \oc / \partial x) \, \delta x$, we can use the properties of the higher moments of $x$.
For thermal motion, and again using the order-by-order approximation\autocite{Brawley2016,Hauer2015}, we use
\begin{equation}
  \varone{\left( x^k \right)^2}_{k \Om} = \frac{k!}{2^{k-1}} \varone{x^2}^k,
\end{equation}
which we can directly substitute into Eq.~\ref{eq:varPfromvarOm} to obtain
\begin{equation}
  \varone{P^2}_{k \Om} = 2 A^2 k! \left( \frac{2 \varone{\delta \omega^2}}{\kappa^2} \right)^k,
  \label{eq:supporderbyorder}
\end{equation}
which we used in the main text.

\begin{figure}
\includegraphics[width=0.5\textwidth]{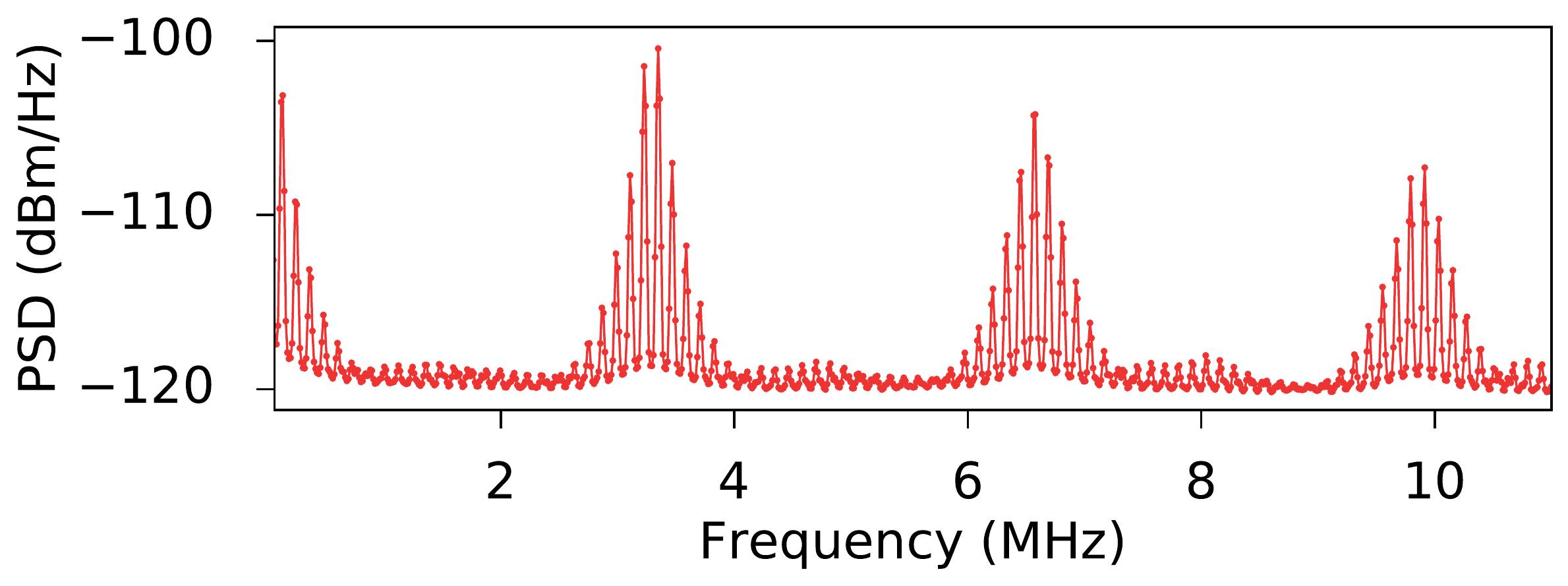} \\
\includegraphics[width=0.6\textwidth]{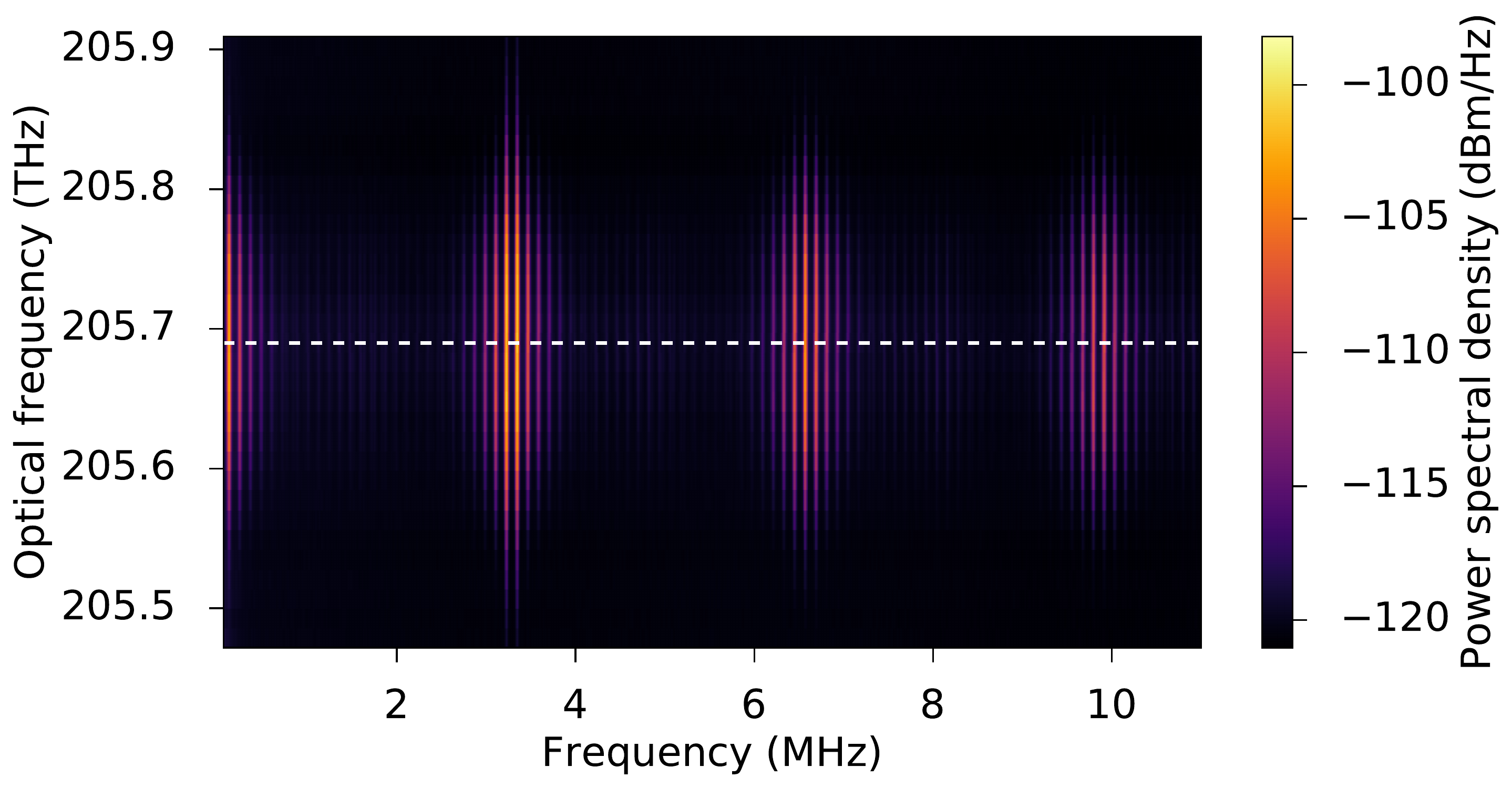}
\caption{ {\bfseries Nonlinear transduction.} Electronic spectrum analyser signal taken at room temperature, as a function of the optical frequency.}
\label{specgram}
\end{figure}

The nonlinear transduction as a function of detuning can be seen in the spectrogram in Fig.~\ref{specgram}.
The crosscut corresponds to Fig.~3a in the main text.
Due to the quadrature-averaging homodyne detection, all peaks show a single-peaked detuning dependence which is maximum at resonance.

\subsection*{Power dependence of transduced mechanical motion}
\begin{figure}
\includegraphics[width=0.5\textwidth]{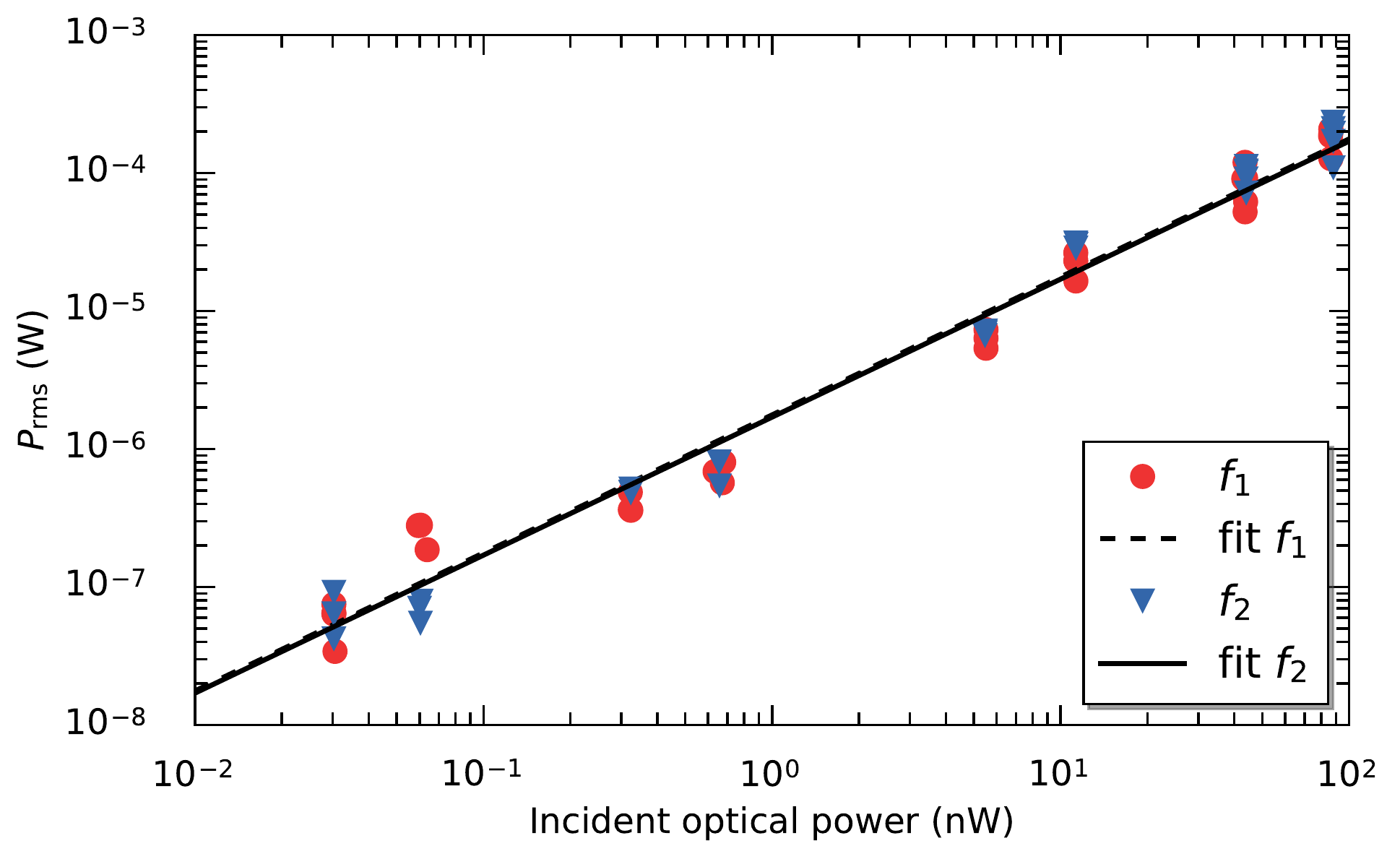}
\caption{ {\bfseries Power dependence.}
Measured band power of the two fundamental mechanical frequencies $f_1$ and $f_2$ as a function of the incident optical power, with the laser frequency on-resonance with the cavity.
}
\label{powerdep}
\end{figure}

The transduction of mechanical motion depends on the optical power used, as shown in the dependence of $\varone{P^2}$ on $A^2 \propto \Pin \Plo$ in Eq.~\ref{eq:supporderbyorder}.
In particular, we expect the measured band power to be linear in both the power in the signal arm $\Pin$, incident on the cavity, and the power in the reference arm $\Plo$.
Deviations from a linear relation indicate heating or cooling of the mechanical motion due to the additional laser power\autocite{Aspelmeyer2014}.
Figure~\ref{powerdep} shows transduced mechanical motion as a function of the incident optical power, with the laser on-resonance with the cavity and while the device was cooled to 3~K.
The data was corrected for the local oscillator power, which was mostly kept constant and did not vary by more than a factor 10.
At higher powers (near 100 nW), fluctuations in the center frequency and shape of the peak could be observed (similar to the effects shown off-resonance in Fig.~\ref{optspring} in the main text), however the peak area still follows the linear dependence as a function of the optical power incident at the structure.
This shows that for all probe powers used here, there is no significant heating due to the incident laser light.

\subsection*{Estimation of coupling efficiency from the optical spring effect}
\begin{figure}
  \includegraphics[width=0.5\textwidth]{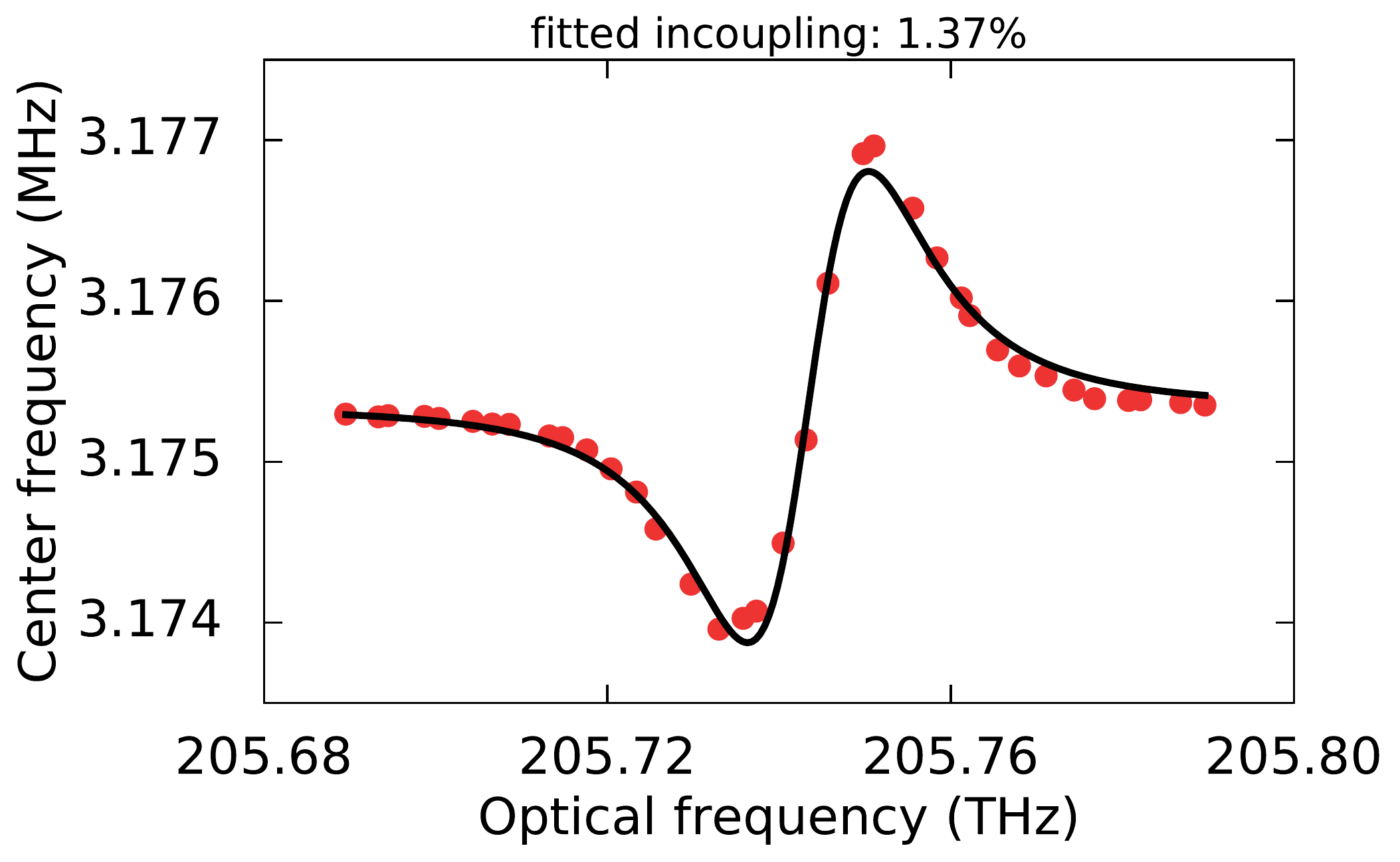}
  \caption{ {\bfseries Optical spring effect at low temperature. }
  Center frequencies of one of the peaks in the mechanical spectra with the structure at 3~K.
}
\label{optspringfit}
\end{figure}

Figure~\ref{optspringfit} shows the fitted center frequency of the mechanical resonance $f_1$ at 3~K with $\Pin = 20.6$~nW, corresponding to the spectrogram shown in Fig.~5a in the main text.
Fitting the change in frequency as a function of the optical detuning with the function describing optical spring effect in the bad-cavity limit\autocite{Leijssen2015,Aspelmeyer2014} provides us with an estimate of the incoupling efficiency.
The average over 5 measurements at different input power yielded a coupling efficiency $\eta=1.3\%$, with a standard deviation of 0.3\%.

This analysis ignores the effects of the large thermal fluctuations, which as we show in the main text broadens and reduces the strength of the optical spring effect. 
This means the obtained coupling efficiency provides a lower limit for the actual coupling efficiency.
However, the numerical model we used to generate the simulated spectra in Fig.~5 in the main text shows that the fluctuations at 3~K mainly broaden the mechanical spectra but don't yet lead to strong reduction in the strength of the optical spring effect.

\subsection*{Radiation pressure force in a cavity with large thermal fluctuations}
We express the radiation pressure force in the cavity as 
\begin{equation}
F_\mathrm{rad} = \frac{\hbar ( \partial \oc / \partial x ) n_c^\mathrm{max}}{1+u^2},
\end{equation}
where $u \equiv\frac{2}{\kappa} (\overline{\Delta} + (\partial \oc / \partial x) x)$ and $n_c^\mathrm{max}$ is the maximum number of photons in the cavity when it is driven at resonance.
We assume the resonator moves harmonically: $x(t) = x_0 \sin \Omega t$.
This results in a time-dependent normalized detuning $u(t) = \overline{u}(\overline{\Delta}) + u_0 \sin \Omega t,$ where $u_0 = 2 (\partial \oc / \partial x) x_0 / \kappa$.

Given an amplitude $x_0$, we extract the first fourier coefficient of the force (at the resulting harmonic frequency $\Omega$):
\begin{equation}
a_1^\mathrm{rad} = \frac{\Omega}{\pi} \int_{0}^{2 \pi / \Omega} F_\mathrm{rad} (t') \sin \Omega t' dt'.
\end{equation}
In analogy with the mechanical restoring force, we can calculate an effective spring constant from the fourier coefficient: with $F = - k x$ and $x(t) = x_0 \sin \Omega t$, $F = \sum_{n=1}^{\infty} a_n \sin n \Omega t$ implies that $a_1 = - k x_0$.

Finally, a correction on the spring constant results in a correction on the frequency $\Omega$:
\begin{equation}
\Omega = \sqrt{\frac{k+k_\mathrm{rad}}{m}} = \sqrt{\frac{k}{m}} \sqrt{1+\frac{k_\mathrm{rad}}{k}},
\end{equation}
which we can approximate using $\sqrt{1+x} \approx 1+x/2+\ldots$:
\begin{equation}
\Omega \approx \sqrt{\frac{k}{m}} (1+\frac{k_\mathrm{rad}}{2k}),
\end{equation}
therefore 
\begin{equation}
  \delta \Omega \approx \sqrt{\frac{k}{m}} \frac{k_\mathrm{rad}}{2k} = \frac{k_\mathrm{rad}}{2 m \Omega}.
\end{equation}

To obtain the results presented in Fig.~\ref{optspring} in the main text, we perform this calculation for a large number of amplitudes $x_0$, sampled from a thermal distribution.
We then average together Lorentzian lineshapes with center frequencies given by the resulting set of frequency shifts.

\subsection*{Characterization of a second device}

\begin{figure}
\includegraphics[width=0.5\textwidth]{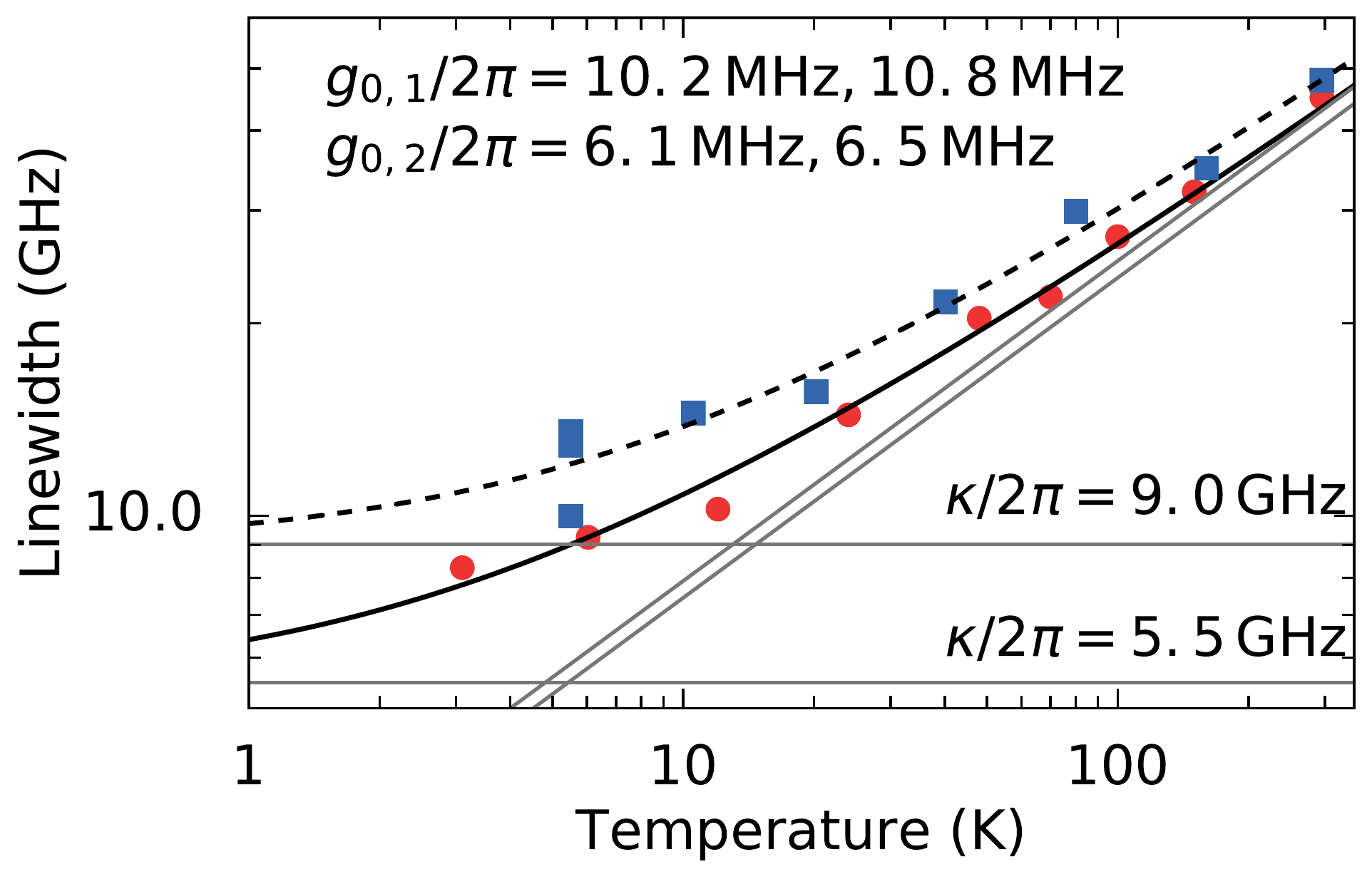}
\caption{ {\bfseries Characterisation of a second device.}
Linewidth versus temperature data, similar to Fig.~\ref{lw}d in the main text.
For the dataset with blue squares, the sample was not shielded from thermal radiation.
}
\label{lwvsT}
\end{figure}

The data labeled as ``device 2'' was taken on a sample with a slightly different design, which did not include the double-period modulation to increase the outcoupling, as well as a different shape of the holes.
The optical linewidth as a function of temperature for this device is shown in Fig.~\ref{lwvsT}, in two separate datasets.
The dataset shown with blue squares was taken with only one window in the cryostat, which was at room temperature, such that the sample was not shielded from thermal radiation.
The other dataset was taken with the second window (cooled to around 10 K) in place.
Therefore the sample temperature for the blue dataset was most likely higher than the thermometer temperature, which we used to plot and fit this data.
Although this leads to a relatively large difference in the fitted optical loss rate $\kappa$, both datasets yield very similar coupling rates $g_0$.

\subsection*{Quadratic displacement measurements}

The signal-to-noise ratio (SNR) of a quadratic displacement measurement, in the regime where the order-by-order approximation is valid, can be found by considering Eq.~\ref{eq:supporderbyorder}, and assuming the measurement is shot-noise limited.
If the optical power in the reference arm is much larger than in the signal arm of the interferometer ($\Plo \gg \Pin$), the power spectral density due to the optical shot noise is given by $S_{PP}^\mathrm{SN} = \hbar \oc \Plo$.
The band power at $2 \Om$ due to thermal mechanical fluctuations is given by Eq.~\ref{eq:supporderbyorder}, with $k = 2$ and $\varone{\delta \omega^2} = 2 \nth g_0^2$, leading to
\begin{equation}
  \varone{P^2}_{2 \Om} = 512 \Pin \Plo \eta^2 \left( \frac{g_0}{\kappa} \right)^4 \nth^2,
\end{equation}
where we also substituted $A^2 = 8 \Pin \Plo \eta^2$.
The power spectral density will show a Lorentzian peak with linewidth $2 \Gamma$ at the frequency $2 \Om$, which means its peak value is related to the band power (area under the peak) as $S_{PP}^\mathrm{max} = \varone{P^2}_{k \Om} / 2 \Gamma$.
Finally, we take the signal-to-noise ratio
\begin{equation}
  \mathrm{SNR} \equiv \frac{S_{PP}^\mathrm{max}}{S_{PP}^\mathrm{SN}} = 256 \frac{\Pin / \hbar \oc}{\Gamma} \eta^2 \left( \frac{g_0}{\kappa} \right)^4 \nth^2
\end{equation}
and set it to 1 to find the minimum detectable average phonon occupation $n_\mathrm{min} \equiv \min(\nth)$, leading to
\begin{equation}
  \frac{1}{\overline{n}_\mathrm{min}} = 16 \left( \frac{g_0}{\kappa} \right)^2 \eta \sqrt{\frac{\Pin / \hbar \oc}{\Gamma}}.
\end{equation}

\begin{figure}
\includegraphics[width=0.5\textwidth]{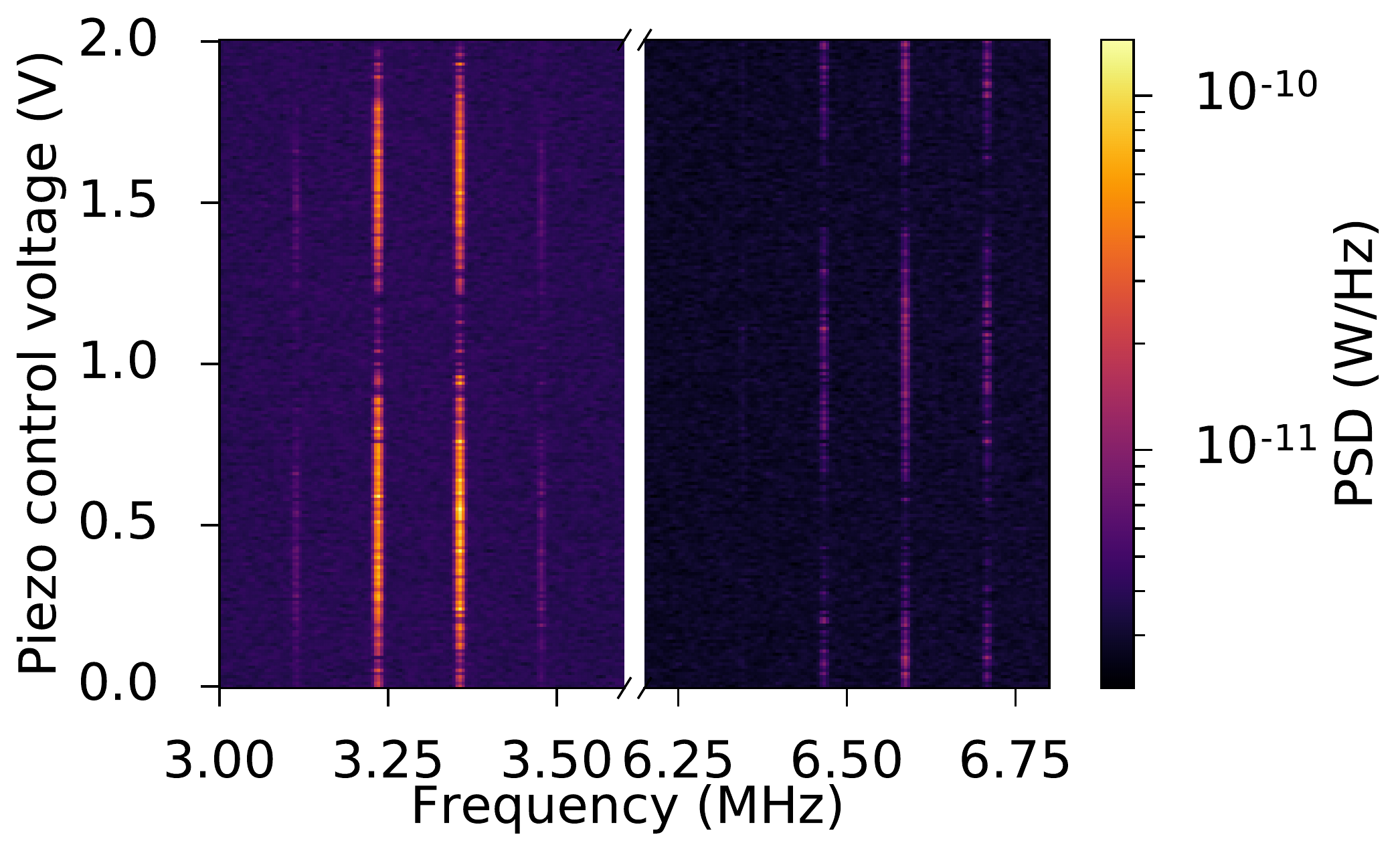}\\
\includegraphics[width=\textwidth]{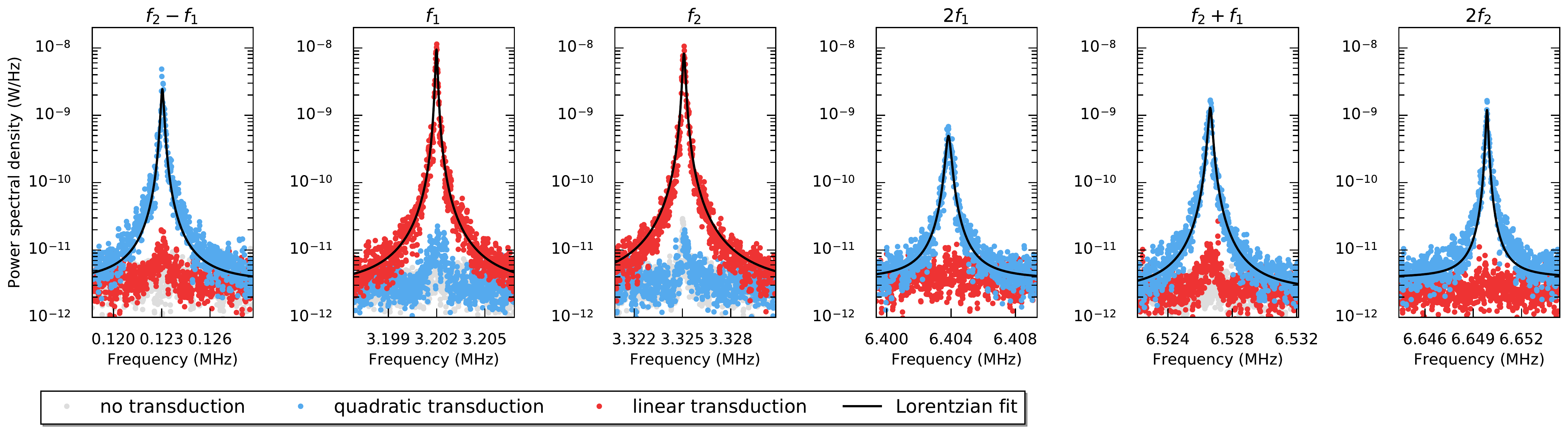}
\caption{ {\bfseries Selective linear and quadratic measurement.} 
Top: spectra showing the first and second order transduction obtained simultaneously, as a function of the piezo mirror position.
Bottom: maximum and minimum linear and quadratic transduction at different settings of the homodyne phase, for all first- and second-order peaks.
Incident optical power was 12.8 nW.
}
\label{piezomirrorsweep_sg}
\end{figure}

Figure~\ref{piezomirrorsweep_sg} shows additional data for the quadratic measurement presented in the main text. 
The top panel shows the peak around the fundamental frequency decrease at the same piezo mirror position where the peaks at twice the frequency are strongest, further confirming the ability to suppress the linear measurement while performing the quadratic measurement.
The bottom panel shows the linear and quadratic measurements at optimum positions of the piezo mirror for both fundamental frequencies and all 4 second-order combinations.

\printbibliography
\end{refsection}
\end{document}